\documentclass[fleqn,usenatbib]{mnras}

\usepackage{newtxtext,newtxmath}
\usepackage[T1]{fontenc}
\usepackage{ae,aecompl}

\usepackage[compact]{titlesec}  
\titlespacing{\section}{0pt}{12pt}{7pt}
\titlespacing{\subsection}{0pt}{10pt}{5pt}
\titlespacing{\subsubsection}{0pt}{8pt}{4pt}
\usepackage{graphicx}	% Including figure files
\usepackage{amsmath}	% Advanced maths commands
\usepackage{soul}
\usepackage[normalem]{ulem}
\usepackage{flushend}
\makeatletter
\renewcommand\sout[1]{\@bsphack\@esphack}%
\makeatother

\title[B1508+55 scintillation]{Scintillation of PSR B1508+55 - the view from a 10000-km baseline}
\author[Marthi et al.]{
V.~R.~Marthi,$^{1,2,3}$\thanks{E-mail:
  \href{mailto:vrmarthi@ncra.tifr.res.in}{vrmarthi@ncra.tifr.res.in} (VRM);
  \href{mailto:dana.simard@astro.caltech.edu}{dana.simard@astro.caltech.edu} (DS);
  \href{mailto:ramain@mpifr-bonn.mpg.de}{ramain@mpifr-bonn.mpg.de} (RAM)}
D.~Simard$^{1,2,4,5}\textcolor{blue}{\footnotemark[1]}$,
R.~A.~Main$^{6,4,1,2}\textcolor{blue}{\footnotemark[1]}$,
U-L.~Pen$^{1,2,7,8}$,
M.~H.~van~Kerkwijk$^{4}$, 
\newauthor K.~Vanderlinde$^{2}$, Y.~Gupta$^{3}$, C.~Roberts$^{9}$ and B.~M.~Quine$^{10,9}$.\\
$^{1}$Canadian Institute for Theoretical Astrophysics, University of Toronto, 60 St. George Street, Toronto, ON M5S 3H8, Canada\\
$^{2}$Dunlap Institute for Astronomy \& Astrophysics, University of Toronto,  50 St. George Street, Toronto, ON M5S 3H4, Canada\\
$^{3}$National Centre for Radio Astrophysics, Tata Institute of Fundamental Research, Post Bag 3, Ganeshkhind, Pune - 411 007, India\\
$^{4}$Department of Astronomy and Astrophysics, University of Toronto,  50
St. George Street, Toronto, ON M5S 3H4, Canada\\
$^{5}$Cahill Centre for Astronomy and Astrophysics, California Institute of
Technology, 1200 E California Drive, Pasadena, CA 91125, USA \\
$^{6}$ Max-Planck Institut fur Radio Astronomie, Auf dem Hugel, Bonn, Germany \\
$^{7}$Canadian Institute for Advanced Research, Program in Cosmology and Gravitation, Toronto, ON M5G 1Z8, Canada\\
$^{8}$Perimeter Institute for Theoretical Physics, 31 Caroline Street North, Waterloo, ON N2L 2Y5, Canada\\
$^{9}$Thoth Technology Inc., Algonquin Radio Observatory, Achray Road,  RR6, Pembroke, ON K8A 6W7, Canada\\
$^{10}$Dept. of Physics and Astronomy, York University, 4700 Keele Street, Toronto, ON M3J 1P3, Canada
}

\date{Accepted XXX. Received YYY; in original form ZZZ}

\pubyear{2020}

\begin{document}
\label{firstpage}
\pagerange{\pageref{firstpage}--\pageref{lastpage}}
\maketitle

\begin{abstract}
We report on the simultaneous Giant Metrewave Radio Telescope (GMRT) and
Algonquin Radio Observatory (ARO) observations at 550-750 MHz of the
scintillation of PSR B1508+55, resulting in a $\sim$10,000-km baseline. This
regime of measurement lies between the shorter few 100-1000~km baselines of
earlier multi-station observations and the much longer earth-space baselines. We measure a scintillation cross-correlation coefficient of $0.22$, offset from zero time lag due to a $\sim 45$~s traversal time of the scintillation
pattern. The scintillation time of 135~s is $3\times$ longer, ruling out
isotropic as well as strictly 1D scattering. Hence, the low cross-correlation
coefficient is indicative of highly anisotropic but 2D scattering. The common
scintillation detected on the baseline is confined to low delays of $\lesssim 1
\mu$s, suggesting that this correlation may not be associated with the parabolic scintillation arc detected at the GMRT. Detection of pulsed echoes and their direct imaging
with the Low Frequency Array (LOFAR) by a different group enable them to measure
a distance of 125~pc to the screen causing these echoes. These previous
measurements, alongside our observations, lead us to propose that there are at least two
scattering screens: the closer 125 pc screen causing the scintillation arc
detected at GMRT, and a screen further beyond causing the scintillation detected on the GMRT-ARO baseline. We advance the hypothesis that the 125-pc screen partially resolves the speckle images on the screen beyond leading to loss of coherence in the scintillation dynamic spectrum, to explain the low cross-correlation coefficient.

\end{abstract}

%%%%%%%%%%%%%%%%%%%%%%%%%%%%%%%%%%%%%%%%%%%%%%%%%%

%%%%%%%%%%%%%%%%% BODY OF PAPER %%%%%%%%%%%%%%%%%%

% This is a simple template for authors to write new MNRAS papers.
% See \texttt{mnras\_sample.tex} for a more complex example, and \texttt{mnras\_guide.tex}
% for a full user guide.

% All papers should start with an Introduction section, which sets the work
% in context, cites relevant earlier studies in the field by \citet{Others2013},
% and describes the problem the authors aim to solve \citep[e.g.][]{Author2012}.

%\section{Methods}

% \begin{itemize}
%     \item intro to scintillation
%     \item current understanding vs new understanding
%     \item sheet model
%     \item brisken et al 2010 work and extensions
%     \item measuring velocities and distances
%     \item multiple screen scattering
%     \item B1508+55 - parallax, distance, PM, echo, etc.
%     \item previous measurement and olaf's measurement(cite)
%     \item our intention in this paper
%     \item organization of this paper
% \end{itemize}

\section{INTRODUCTION}\label{sec:intro}

Interstellar scintillation (ISS) of pulsars has been used as a tool to study the
ionized interstellar medium (IISM) and as a probe of pulsar
magnetospheres. Pulsars, being unresolved at all radio wavelengths and emitting
coherent radiation, are the best sources to study the IISM through their
scintillation. Scintillation arises from constructive and destructive interference of the
scattered rays, causing modulation of intensity in the time-frequency plane
\cite[see e.g.][]{Cordes2006}. Pulsar dynamic spectra often exhibit rich features caused by scintillation. The
Fourier conjugate of the dynamic spectrum $I(t,\nu)$, where $t$ is time and
$\nu$ is frequency, is called the conjugate
spectrum, squaring which gives the secondary spectrum $S(f_\mathrm{D}, \tau) =
|\tilde{I}(f_\mathrm{D}, \tau)|^2$, where $f_\mathrm{D}$ is the differential
Doppler frequency and $\tau$ is the differential delay. In simple words, the
secondary spectrum is the two-dimensional power spectrum of the intensity dynamic spectrum.

There has been considerable evolution in our understanding of
interstellar scintillation in the last two decades. Our currently held view owes its beginnings to the
discovery of parabolic scintillation arcs \citep{Stinebring2001} in the secondary spectra of pulsars, challenging the notion of a largely volume-filling
medium that had been posited as the origin for scattering. The presence of thin,
well-defined, parabolic features in the secondary spectra of several
pulsars \citep{Hill2003, Cordes2006} means that highly anisotropic scattering occurs in localized
regions along the line of sight. Evidence for
anisotropic scattering came from direct mapping, by \cite{Brisken2010}, of the scattered images on the
sky aided by the resolving power of scintillation, far exceeding what is
possible through conventional VLBI: they found a densely
packed series of point-like scattered images of the pulsar, aligned roughly
along a straight line, with the magnifications and number of speckles tapering
off on either side of the line of sight. The
separations between the speckle images on the screen range between $\sim0.1$ and
$\sim10$ AU, the scales at which irregularities are thought to exist on the
screen. These speckle images hint at a filamentary structure or structures in the IISM
as the locales for the anisotropically scattered images.

In a scenario involving roughly spherical lenses, the high electron
overdensities would translate into physically untenable overpressured regions,
ruling out a volume-filling turbulent medium. To circumvent this difficulty, \citet{Pen2014b} advance a model where they invoke very thin
plasma reconnection sheets confined between regions of misaligned magnetic
fields. Perturbations propagate as Alfv\`{e}n-like surface waves, the amplitudes of which are
very small in relation to the extent of the sheet. These sheets extend for
several hundred AU but are highly inclined with the line of sight, causing
extremely high projected local overdensities by virtue of which they have very
high lensing potential. This model convincingly explains the observed
anisotropic scattering reported by \citet{Brisken2010}. If these structures
persist in the sheets over long periods of time (compared to typical observing durations), as suggested by the
\citealt{Pen2014b} model, the images would appear to move relative to the
pulsar due to the relative motion of the pulsar and the reconnection
sheet. Earlier work by \cite{Hill2005} indeed measured uniform motion of the
arclets along the main parabola suggesting that these features remain
practically stationary in the screen as the pulsar moves behind
it. \cite{Simard2018} further analyze this model quantitatively to study the
expected motion of the lensed images, effectively predicting their positions and
magnifications in future time and out-of-band frequencies. Lensed images in the
reconnection sheet model are caused by wave crests, each of which can be
described by a single parameter. This facilitates a completely deterministic
treatment of each lensed image, allowing a posteriori prediction of
scintillation.

\begin{figure*}
    \centering
    \includegraphics[scale=0.46]{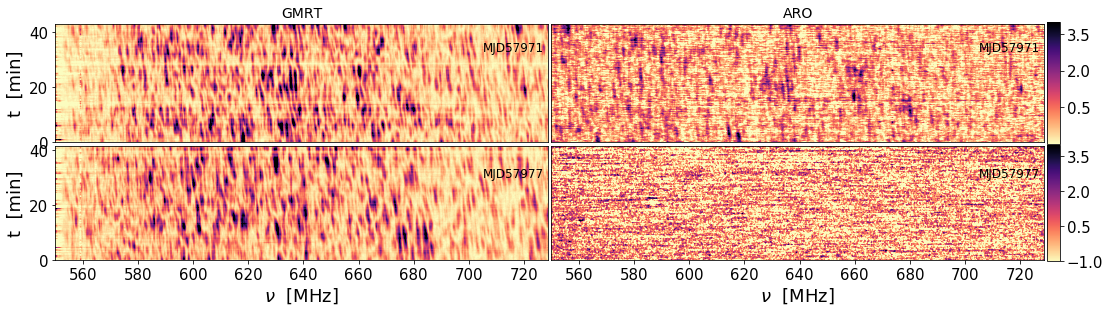}
    \caption{The dynamic spectra for  B1508+55 from the two
      observing sessions at GMRT and ARO are shown here, between 550 and 730
      MHz. Each pixel is $15P_0 \times 293$ kHz. The colour scale in all the
      dynamic spectra ranges -1 to 4. The S/N of the ARO dynamic
      spectrum on MJD57977 is very poor because of a tracking malfunction. Only
      the first $\sim40$ minute scan from the GMRT on MJD57977 is shown here.}
    \label{fig:GMRT-dynamic-spectra}
\end{figure*}

\subsection{PSR B1508+55}
The pulsar we study in this paper is B1508+55, which 
has among the highest measured proper motion velocities, moving at
$963^{+61}_{-64}$ km~s$^{-1}$ south east and a measured parallax distance of
$2.1\pm0.1$~kpc \citep{Chatterjee2009}. Earlier observations of this
pulsar \citep{Stinebring2007} show thick horizontal streak-like features instead of parabolic reverse
arclets in the secondary spectrum, besides straggler arclets that are not strictly aligned with the
parabolic locus. An extra pulse component was initially detected with the German 
Long Wavelength Consortium (GLOW) telescopes (Oslowski et al., \emph{in 
prep.}). Follow-up observations revealed that this component is 
approaching the main pulse, subsequently identified to be likely of an ISM 
origin\footnote{S. Os\l{}owski is running a B1508+55 monitoring
    campaign with LOFAR; private communication}. Further LOFAR observations have now revealed the
  presence of multiple images \citep{Wucknitz2019} roughly aligned along the proper
  motion vector, suggesting that the pulsar is moving towards perhaps linearly
  aligned refracting structures. \citet{Bansal2020} report the detection of these
  echoes at frequencies below 100~MHz.

Given the peculiar thick-arc nature of the secondary spectrum and the imaging of the echoes, the two main goals in this work are: (i) to measure the scintillation time and the scintillation pattern delay on an orthogonal pair of long baselines at 650 MHz using dual station simultaneous observations in the 550-750~MHz band and (ii) to
measure the correlation coefficient along the baselines and infer if the scattered image is one- or two-dimensional.

\subsection{Thin screen scattering}\label{ssec:defn}
It is useful to have a brief perspective of localized thin screen scattering to
aid in some of the discussion in later sections. Definitions of quantities used throughout the remainder of this paper are in order: if $s$ is the
fractional distance from the pulsar at which the screen is located, i.e. $s
= 1 - d_\mathrm{scr}/d_\mathrm{psr}$, we define the effective distance and effective
velocity as

\begin{equation}
    D_\mathrm{eff} = \frac{1-s}{s} d_\mathrm{psr}
    \label{eqn:D_eff}
\end{equation}
and
\begin{equation}
    \mathbf{V}_\mathrm{eff} = \frac{1-s}{s} \mathbf{V}_\mathrm{psr} + \mathbf{V}_\mathrm{obs} - \frac{1}{s}\mathbf{V}_\mathrm{scr}
    \label{eqn:V_eff}
\end{equation}
where $d_\mathrm{psr}$ is the distance to the pulsar from the observer and
$d_\mathrm{scr}$ is the distance to the scattering screen. $\mathbf{V}_\mathrm{psr}$,
$\mathbf{V}_\mathrm{obs}$ and $\mathbf{V}_\mathrm{scr}$ are the velocities of the
pulsar, observer and the screen respectively. The equations and the
formalism presented here apply strictly to one-dimensional (1D) scattering.

Since the scattered ray suffers an additional path
length, it picks up a differential time delay $\tau$
proportional to the square of the angular displacement from the line of sight,
$\theta$ \citep{Stinebring2001, Hill2003, Cordes2006}:
\begin{equation}
    \tau = \frac{D_\mathrm{eff}}{2c} \theta^2.
    \label{eqn:tau}
\end{equation}
Due to the uniform motion of
the pulsar and the stationarity of the images on the screen (assuming that the
lifetime of substructure on the screen is much longer than an observation), the differential delay changes
at a rate proportional to the effective velocity, i.e. linear in angular separation between pairs of images,
\begin{equation}
    f_\mathrm{D} = \nu \frac{d\tau}{dt} = \btheta \cdot \mathbf{V}_\mathrm{eff}/\lambda
    \label{eqn:f_D}
\end{equation}
from which it follows that
\begin{equation}
    \tau = \eta f_\mathrm{D}^2,
    \label{eqn:quadratic}
\end{equation}
where $f_\mathrm{D}$ is the differential
Doppler frequency and $\eta$ is the curvature of the parabolic arc:
\begin{equation}
    \eta = \frac{\lambda^2}{2c}\ \!
    \frac{D_\mathrm{eff}}{|\mathbf{V}_\mathrm{eff}|^2\ \! \mathrm{cos}^2 \alpha_s}.
    \label{eqn:curvature}
\end{equation}
$\alpha_s$ is the angle that the line of scattered images, which we henceforth
call the scattering axis, makes with the effective velocity vector
$\mathbf{V}_\mathrm{eff}$. The scintillation pattern, therefore, always moves
along the scattering axis opposite to the direction of the projected effective
velocity vector as seen by an observer on Earth.
The curvature is degenerate in the effective distance $D_\mathrm{eff}$ and the
effective velocity $\mathbf{V}_\mathrm{eff}$, precluding an independent measurement of
both the quantities without explicit or implicit assumptions. Single station
measurements of the curvature are therefore generally used with approximate
effective velocities to estimate screen distances.  A common assumption is that
the largest contribution to the effective velocity accrues from the pulsar's
proper motion velocity \citep[see e.g.][]{Stinebring2001}.
Using a pair of radio telescopes, the apparent speed
of the scintillation pattern along that baseline,
$V^{\mathrm{app}}_\mathrm{ISS}$, can be measured. An independent measurement of the full 2D
velocity of the scintillation pattern, measured using scintillation pattern
traversal times between multiple stations observing simultaneously, would hence break the
degeneracy between $D_\mathrm{eff}$ and $\mathbf{V}_\mathrm{eff}$.

\section{OBSERVATIONS AND DATA}\label{sec:observations}
\subsection{Simultaneous wideband observations}\label{ssec:wideband}
Observations of  B1508+55 were carried out in two sessions in August 2017
simultaneously with the Giant Metrewave Radio Telescope (GMRT) and the Algonquin
Radio Observatory (ARO) 46-m telescope. The GMRT observations were made with the 500-850
MHz Band-4 wideband feed, which were deployed on 15 antennas at the time. These feeds
are now available on all 30 antennas and comprise one of the six wavebands of
the upgraded GMRT (uGMRT; \citealt{Gupta2017}). We used the GPU-based GMRT Wideband Backend (GWB; \citealt{Reddy2017}) to
process 550-750 MHz of Band-4. For the observations we report here, the GWB was
deployed as a phased-array beamformer, in which
voltages from the antennas are summed in phase before detection and
sub-integration. The GWB was configured to split the 200-MHz bandwidth into 8192 channels to give
a frequency resolution of 24.4 kHz, and a time resolution of 262.144 $\mu$s.

The ARO 46m dish, although not primarily an
astronomical facility at present, is equipped with a 400-800 MHz feed designed
and mass produced for the Canadian Hydrogen Intensity Mapping Experiment
(CHIME). The entire band is digitized utilizing the bandpass sampling technique
and recorded as baseband voltages with 1024 channels, implemented using a polyphase filter
bank \citep{Recnik2015}.

We chose to observe on two different days, at LSTs
separated by 6 hours, to realize a pair of orthogonal baselines aided by the high declination of the pulsar. It is sufficient
that the scattering geometry is similar on both of the days. The exigent scheduling constraints at the GMRT meant that the closest such pair of observing sessions would be spaced $\sim150$ hours apart. 

In the first observing session on 2017-August-06
(MJD57971), B1508+55 was observed approximately for 45 minutes with the GMRT
(of the one hour allocated on the day) and ARO simultaneously when the source
was at an hour angle of -5h at the GMRT. The pulsar is circumpolar at
ARO.

Observations of the second session on 2017-August-12 (MJD57977) commenced at a
local sidereal time (LST) 6 hours later than on the first day and continued for
3 hours at both telescopes. Unfortunately the ARO 46m drive system malfunctioned
and mistracked the source co-ordinates, rendering the data entirely unusable. However, at the
GMRT, observations were carried out for the full 3 hours allocated, resulting in
a total of 3 $\sim40$-minute scans. Only the first $\sim40$-minute scan
is shown in Figure~\ref{fig:GMRT-dynamic-spectra}. The details of the observations
are summarized in Table~\ref{tab:observations}.

\begin{table}
	\centering
	\caption{Details of the observations of  B1508+55 with the GMRT and ARO. $P_0$ is the period of B1508+55.}
	\label{tab:observations}
	\begin{tabular}{l|l|l} % four columns, alignment for each
		 \hline
		\textbf{Setup} &  \textbf{GMRT} & \textbf{ARO} \\
		\hline
		Band(MHz) & 550-750 & 400-800\\
		\textbf{Scan start} & & \\
		MJD57971 & 8:24:50 UTC & 7:30:25 UTC \\
		MJD57977 & 14:20:00 UTC & 14:00:00 UTC \\
		\textbf{Scan stop} & & \\
		MJD57971 & 9:10:02 UTC & 9:41:40 UTC \\
		MJD57977 & 17:31:21 UTC & 18:05:00 UTC \\
		Number of channels & 8192 & 1024 \\
		Native time resolution & 262.144 $\mu$s & 2.56 $\mu$s \\
		Final time resolution ($t_\mathrm{s}$) & 262.144 $\mu$s & $P_0/128$ \\
		Mode & Phased array intensity & Baseband voltage \\
	\hline
	\end{tabular}
\end{table}

\subsection{Data reduction}\label{ssec:reduction}

The channelized GMRT data were dedispersed with the DM of 19.6191 pc cm$^{-3}$.
Next, the band-averaged
data were folded at the the period of the pulsar over $P_0/t_\mathrm{s}$ bins
to determine the ``on'' phase. Then, for each single pulse for every channel,
the mean of the ``off'' pulse was subtracted and the result divided again by the
mean ``off'' pulse, to establish bandpass correction and to obtain the dynamic
spectrum. Finally, the dynamic spectrum was normalized by dividing by its mean
and subtracting unity. This final data product is used for computing the
modulation index.

The ARO dynamic spectrum was similarly obtained after sub-integrating the pulse to 128 phase bins
across the period. Since B1508+55 exhibits large pulse-to-pulse flux
variability, we visually aligned the band-averaged pulse series to determine the
clock offset between GMRT and ARO on MJD57971.

\section{RESULTS}
\subsection{Dynamic and Secondary Spectra}\label{ssec:dspec-and-sspec}
The dynamic spectra of the GMRT and ARO from both sessions of the observations
of  B1508+55 are shown in Figure~\ref{fig:GMRT-dynamic-spectra}. Frequencies
above 730~MHz are excluded from our analysis owing to severe mobile
communication RFI at ARO. The autocorrelation of the normalized dynamic spectrum at zero time and frequency lag is hence the modulation index squared $m^2$ by
definition \citep{Johnson2012}:
\begin{equation}
  m^2 = \frac{\langle (I-I_n)^2 \rangle}{\langle I - I_n \rangle^2} - 1
  \label{eqn:mindex}
\end{equation}
where $I$ is the intensity time series and $I_n$ is the `off-pulse' intensity
time series which contains the receiver noise.

\begin{figure}
    \centering
    \includegraphics[scale=0.31]{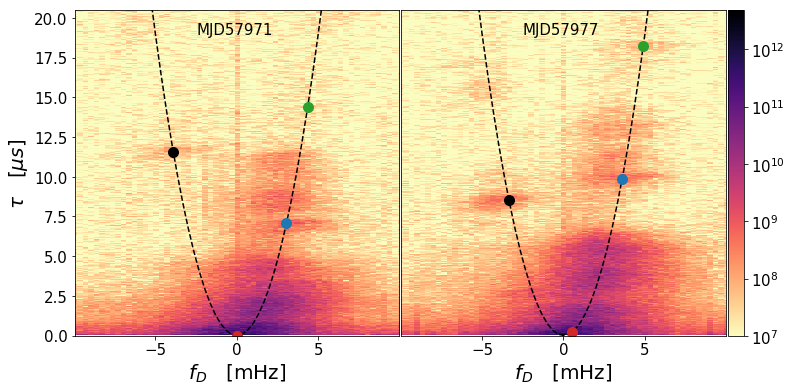}
    \caption{GMRT secondary spectra of  B1508+55 on MJD57971 and
      MJD57977, obtained from the dynamic spectra shown in Figure~\ref{fig:GMRT-dynamic-spectra}. The markers indicate the positions of the features that have
      moved by an amount $\Delta f_D = 0.546$ mHz, from which the curvature
      of the parabola $\eta = 0.76$ s$^3$ is estimated.}
    \label{fig:GMRT-sec-spectra}
\end{figure}

Due to their superior S/N, we will restrict our discussion of the
secondary spectra to the GMRT observations. The secondary spectra for MJD57971 and
MJD57977 are both shown in Figure~\ref{fig:GMRT-sec-spectra}. The native frequency
resolution of the dynamic spectra is $\sim24.4$ kHz; the highest delay in the
secondary spectrum is thus 20.48 $\mu$s. There are several salient features to
note: on both days, a substantial fraction of the power in scintillation
is confined to $\lesssim$ 5 $\mu$s; at smaller delays, the lack of a clearly
discernible parabolic feature complicates estimating the curvature. A
striking absence of clear, well-defined reverse arclets, useful in identifying
apexes through which to fit the main parabola, exacerbates the problem. The most
notable difference between the two days is the systematic
shift in the features identified with markers. 

A simple approximation for the total shift in $f_\mathrm{D}$, valid when the
curvature of the putative parabola has undergone  no change, is given by
\begin{equation}
    \Delta f_\mathrm{D} = \frac{\Delta t}{2\eta\nu}
    \label{eqn:delay-rate-change}
\end{equation}
which allows us to track isolated features along the main parabola in the
secondary spectrum that correspond to interference between the lensed images and
the line-of-sight ray from the pulsar as it moves behind the screen. Features in
the parabola always move from negative through zero to positive Doppler frequencies: the
delay decreases on the head-side and increases on the tail-side as the pulsar
moves with respect to the stationary images.

 We use different trial values for $\eta$ in eqn.~\ref{eqn:delay-rate-change} to
determine the total shift in the differential Doppler frequency ($\Delta
f_\mathrm{D}$). We identify the curvature ($\eta = 0.76$ s$^3$) that best predicts the net motion of the isolated
arclet-like features. The markers identify those features that moved $\Delta f_D
= 0.546$ mHz between the two observing sessions. Similar motion of apexes of
reverse arclets has been measured empirically by \cite{Hill2005} for 
B0834+06.

 The qualitative similarities with earlier observations of this pulsar \citep{Stinebring2007},
 such as the thick horizontal streak-like features instead of reverse arclets, as well as
 stragglers that stray from the parabolic locus, are
 unmistakable. \cite{Gwinn2019a} discusses the effects of wideband observations
 on the secondary spectrum: averaging over a large band causes a continuum shift
 in the curvature, resulting in thick streak-like features along the differential
 Doppler frequency. Wideband pulse intensity  modulation could as well lead to smearing of features along thee Doppler frequency axis. However, we have confirmed that neither is the case with
 our observations of B1508+55 by a few independent means: (i) we weighted the
 dynamic spectrum by a two-dimensional Hanning window to obtain the secondary
 spectrum, (ii) we binned the dynamic spectrum 60$\times$ in time (i.e. 60$P_0$), (iii) we inspected the secondary spectra obtained over a much narrower subband ($\sim25$ MHz), and
 (iv) we corrected for the wide bandwidth by linearly scaling the dynamic spectrum along
 the frequency axis, normalized for the reference frequency, followed by a
 discrete Fourier transform, to concentrate the power on the reference
 parabola. The horizontal streak-like features persisted in all the above cases, suggesting that this is an intrinsic feature of the scintillation of  B1508+55 at these frequencies.
 
 \subsection{Cross-correlation of scintillation}
\begin{figure*}
    \centering
    \includegraphics[scale=0.5]{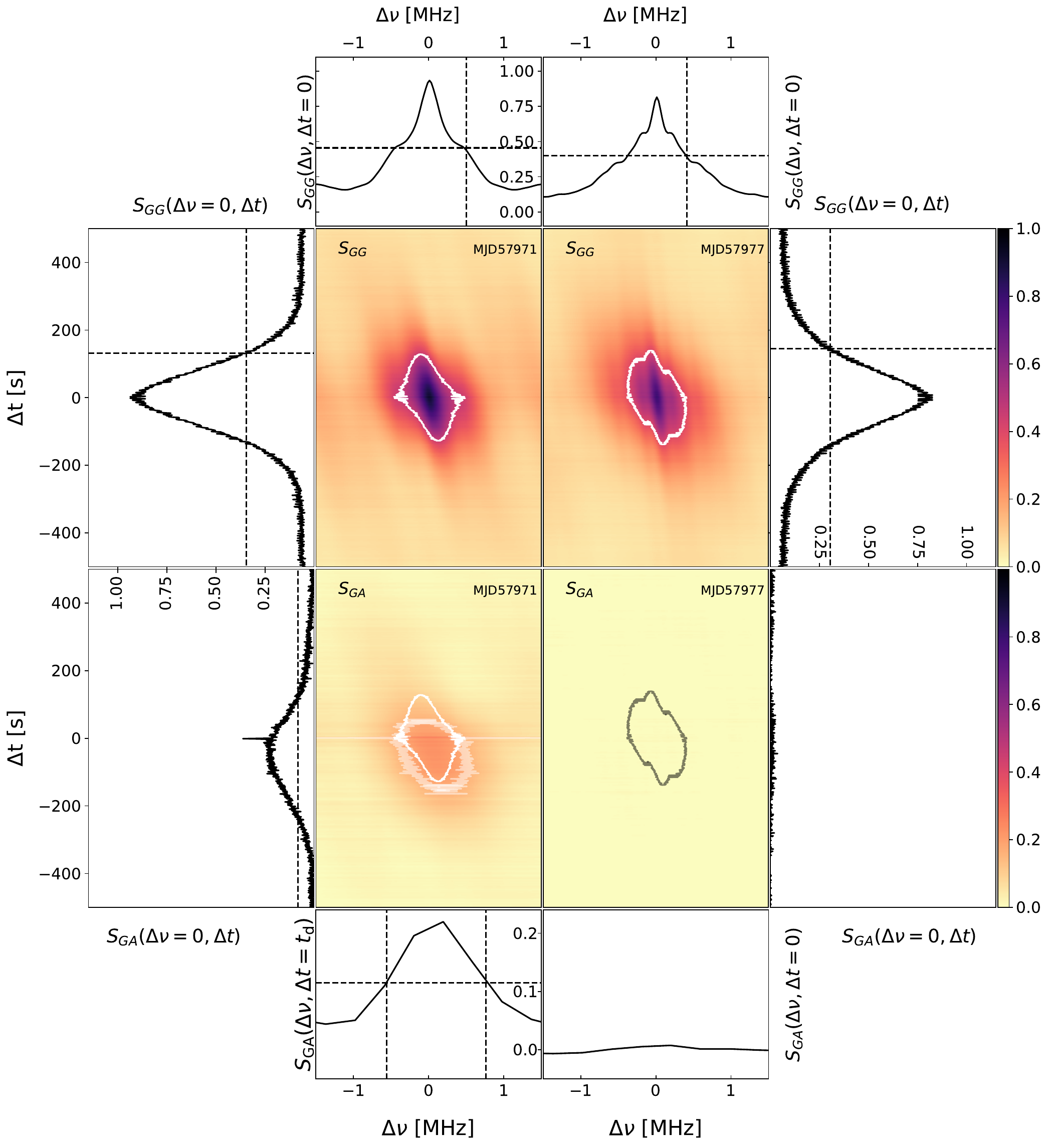}
    \caption{Autocorrelation of scintillation dynamic spectra of GMRT and
      GMRT-ARO cross-correlation of scintillation dynamic spectra. The top row
      of the 2D panels shows respectively the autocorrelation of the GMRT dynamic
      spectra $S_\mathrm{GG}$ on MJD57971 and MJD57977. The bottom row of the 2D
      panel shows the corresponding cross-correlation of GMRT-ARO scintillation
      dynamic spectra, $S_\mathrm{GA}$. The white contours in the top row are at half-maximum of
      the respective autocorrelation peaks while they are overlaid on the
      cross-correlations in the bottom 2D panel. The peak of the
      cross-correlation is offset $\sim$ -45s from zero lag in time. The peripheral plots are cuts of the auto- and
      cross-correlations through $\Delta\nu$ and $\Delta t$ where the 2D function peaks, from which we measure the
      scintillation bandwidth (half-width at half-maximum) and scintillation
      time (half-width at 1/$e$) listed in
      Table~\ref{table:measured_parameters}. }
      \label{fig:6-vs-12-cross-corr}
\end{figure*}

Each pixel of the raw GMRT dynamic spectrum is \mbox{$P_0 \times 200/8192$~MHz}, and of
the ARO dynamic spectrum is \mbox{$P_0 \times 400/1024$~MHz}. We therefore bin the GMRT
dynamic spectrum to the same resolution as that of ARO ($16\times$). 
One of the dynamic spectra is shifted along the time axis to reverse the relative clock-offset
between the observatories.
The dynamic spectra, each with 1024 channels across the 550-750~MHz band, are then
transformed to their respective conjugate spectra and cross-multiplied according
to~(\ref{eqn:cross-secondary-spectrum}), and transformed back to give the
real-valued cross-correlation function (CCF) between the pair of dynamic spectra. The
autocorrelation function (ACF) of the GMRT dynamic spectra are similarly obtained, but retaining the native frequency resolution. Figure~\ref{fig:6-vs-12-cross-corr} shows the 2D GMRT ACF
and 2D GMRT-ARO CCF, along with the time- and frequency-lag cuts
in the peripheral panels. The ARO ACFs are much noisier due to the significantly poorer S/N of the dynamic spectra, and are not shown here.

\begin{figure}
    \centering
    \includegraphics[scale=0.42, trim=0.25cm 0.25cm 0cm 0cm, clip=true]{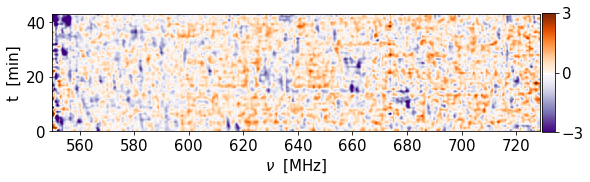}
    \caption{The GMRT-ARO difference dynamic spectrum for  B1508+55 on
      MJD57971. The normalized ARO dynamic spectrum, shifted by the measured
      scintillation delay, has been subtracted from the
      normalized GMRT dynamic spectrum, showing the residual scintillation at
      GMRT and ARO.}
    \label{fig:difference-dynamic-spectra}
\end{figure}

We apply the arguments of \cite{Johnson2012} to determine the true correlation
coefficient from the modulation index squared. The modulation index at zero temporal and spectral lag has contributions from the intrinsic pulse variability and noise. We therefore interpolate the values at
small non-zero spectral and temporal lags through zero lag to determine the correlation
coefficient. Note that the autocorrelation coefficient thus obtained from non-zero temporal lags is
free from the inherent bias introduced by intrinsic pulse variability \citep{Johnson2012}. The
cross-correlation coefficient of the dynamic spectra from GMRT and ARO at zero
temporal lag has the pulse variability contribution, but the peak of the cross-correlation shifts to a non-zero temporal lag
due to the delay in the arrival of the scintillation pattern in one of the
telescopes.

The scatter broadening
($\tau_\mathrm{sc}$) expected from the measured scintillation bandwidth is $\lesssim1\mu$s,
which is much smaller than the $\sim2.6$ms time resolution of the recorded beam. We notice two conspicuous differences between the two GMRT 2D ACFs: first, the
peak feature is narrower in frequency lag and, second, a small but perceptible rotation of the major
axis on the second day with respect to the first. The structure in the frequency-lag ACF is evidence for multiple fringe
spacings, corresponding to the many distinct islands of power in the GMRT
secondary spectra shown in Figure~\ref{fig:GMRT-sec-spectra}.

The normalized CCF peak on MJD57971 is $\rho=0.22\pm0.01$, which is in the same
units as the ACF ($m^2 = 0.91\pm0.01$), allowing for a meaningful and
self-consistent comparison. There is no detectable correlation on
MJD57977. The cross-correlation of $\sim 20 \%$ is a crucial and a
significant measurement. This is not entirely unexpected as, even visually, the dynamic
spectra look different between GMRT and ARO on MJD57971 in
Figure~\ref{fig:GMRT-dynamic-spectra}.  Figure~\ref{fig:difference-dynamic-spectra} shows  the
presence of residual scintillation at both ARO and GMRT, which underpins the low
cross-correlation coefficient.

\subsection{Scintillation delay on the baseline}\label{ssec:scintillation-velocity}

We notice that the peak of the cross-correlation in
  Figure~\ref{fig:6-vs-12-cross-corr} is offset from zero-lag. This lag can also be
  robustly estimated by measuring the corresponding phase gradient in the Fourier plane. 

For dynamic spectra $I_{\mathrm
  G}(\nu, t)$ and $I_{\mathrm A}(\nu, t)$ observed at the GMRT and ARO
respectively, we obtain the quantity
\begin{equation}
\tilde{S}_\mathrm{GA}(\tau, f_\mathrm{D}, \mathbf{b}) = \tilde{I}_\mathrm{G}(\tau, f_{\mathrm D}) \times \tilde{I}^\ast_{\mathrm A}(\tau, f_{\mathrm D}),
\label{eqn:cross-secondary-spectrum}
\end{equation}
which \citet{Simard2019a} call the intensity cross secondary spectrum, where
$\tilde{I}$, the 2D Fourier Transform (2DFT) of $I$, is called the conjugate
spectrum \citep{Brisken2010}. 

\begin{figure}
  \centering
  \includegraphics[scale=0.32]{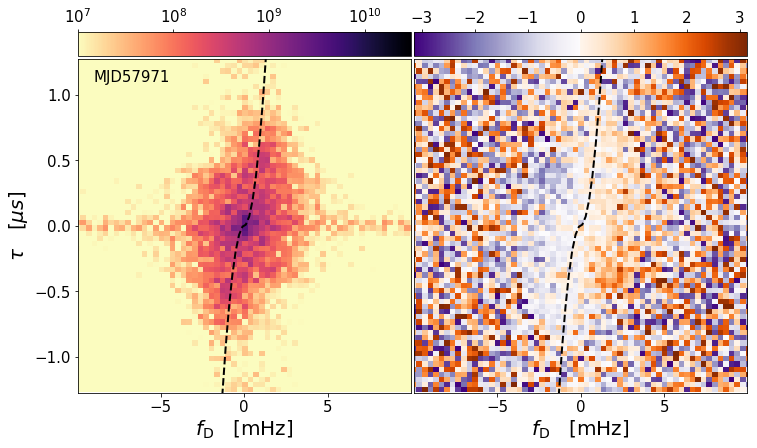}
     \caption{The log magnitude and phase (in radians) of the cross secondary spectrum of scintillation
       of  B1508+55 on MJD57971. The parabola
$\tau= \mathrm{sgn}(f_\mathrm{D})\ \! \eta\ \! f_\mathrm{D}^2$ for $\eta=0.76$
       s$^3$ is overlaid on the cross secondary spectral magnitude and phase.}
     \label{fig:6-vs-12-cross-spec}
\end{figure}

Figure~\ref{fig:6-vs-12-cross-spec} shows the magnitude and phase of the cross
secondary spectra. We detect a phase gradient at $|\tau| \lesssim 1.0 \mu s$ on MJD57971 and none on
MJD57977, the latter due to a telescope malfunction at ARO.

The phase gradient $d\Phi/df_\mathrm{D}$ corresponds to a time delay in the
arrival of the scintillation pattern in one telescope with respect to the other:
\begin{equation}
t_\mathrm{d} = \frac{1}{2\pi}\frac{d\Phi}{df_\mathrm{D}}
\end{equation}

\begin{figure}
    \centering
    \includegraphics[scale=0.57]{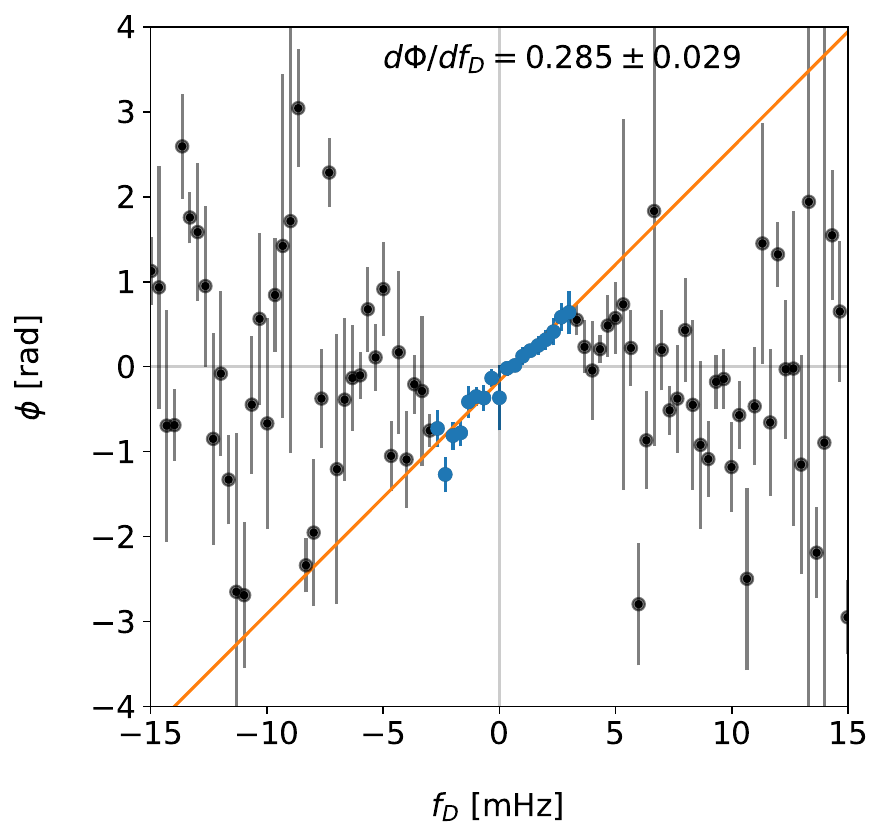}
    \caption{The delay-averaged phases for $\tau\!\!>\!\!0\ \!\mu$s against the differential Doppler
      frequency obtained from the cross secondary spectrum on MJD57971. The
      black points represent the measurement, while only the blue points have been included
      for determining the fit, shown by the solid orange line.}
    \label{fig:fD-vs-phi}
\end{figure}

Figure~\ref{fig:fD-vs-phi} shows the delay-averaged phase across Doppler frequency
measured on MJD57971. The errors on each point have been propagated from the standard deviation of the
real and imaginary parts. The solid line is the inverse noise variance weighted fit to the gradient, the slope of which is
$d\Phi/df_\mathrm{D}=0.285\pm0.029$ rad~mHz$^{-1}$. The error derived
from the covariance matrix of the fit parameters is only 0.016 rad mHz$^{-1}$. The small error is not
surprising given the tiny error bars on the points in blue. However, we expect
the errors to be slightly higher owing to uncalibrated systematics: a 
conservative error of 10\% is reasonable. This gradient translates into a
scintillation pattern arrival delay of $t_\mathrm{d} = 45.3 \pm 4.5$ s, the pattern arriving 
 earlier at the GMRT.
Table~\ref{table:measured_parameters} additionally lists the baseline vectors,
pulsar proper motion and Earth velocity vectors and their relative
orientations. The apparent speed of scintillation along the baseline is thus
$220\pm22\ \mathrm{km\ \!s^{-1}}$, which is an upper limit to the true
sicntillation pattern velocity in the absence of a full 2D measurement of the
scintillation pattern motion.

\begin{table}
	\centering
	\caption{Measured parameters. The apparent scintillation speed measured
          on the baseline is negative as it is opposite in direction to the
          pulsar's proper motion and hence the effective velocity. The baseline
          and velocity components are along $\mathbf{u}$ and $\mathbf{v}$ directions.}
	\label{table:measured_parameters}
	\begin{tabular}{l|c|c} 
	    \hline
		 Parameter & MJD57971 & MJD57977 \\
		 \hline
		 Correlation coefficient ($m^2$) & $0.91\pm0.01$ & $0.80\pm0.01$ \\
		 Scintillation bandwidth (kHz) & 500 & 410 \\
		 Scintillation time (s) & $132\pm3$ & $145\pm3$ \\
		 \hline
		 Baseline length (km) & 9974.3 & 10468.0 \\
		 Baseline vector ($\times 1000$ km) & (-4.2, -9.0) & (9.7, -4.0)\\
		 \hline
                 Pulsar PM angle with BL & 25$^\circ$ & 120$^\circ$ \\
         PM component along baseline (km s$^{-1}$) & 876 & -437 \\
         \hline
		 Earth velocity vector (km s$^{-1}$) & (2.7, 27.9) & (-0.3, 28.2) \\
		 Component along baseline (km s$^{-1}$) & -26 & -11 \\
		 \hline
                 Apparent scintillation speed (km s$^{-1}$) & $-220 \pm22$ & NA \\
                 along baseline & & \\
                 \hline
	\end{tabular}
\end{table}

\section{ASSUMPTIONS FOR A SIMPLE INTERPRETATION}

\subsection{Arc curvature}
Our first assumption is that the curvature of the forward parabola does not
change between the two epochs of observations, as they are only separated by 150
hours. The extrema of annual modulation of scintillation occurs six months
apart. As a fraction of this duration, 150 hours is only 3.4\%. 
The curvature of the forward parabola is measured to be $\eta =
0.76\ \mathrm{s}^3$. While absence of clearly marked reverse arclets precludes
fitting a parabola through their apexes, the best fit curvature parameter is
estimated from the reflex motion of the thick arclet-like features. Besides, we
rely on identifying the same arclet feature to measure the net displacement along
the Doppler axis, which constrains how far apart such measurements are made. For
a close pair of adjacent measurements, one can hence always assume $d\eta/dt=0$ and
still reliably measure the annual modulation of the curvature with regularly
spaced observation pairs. The full equation, however, is easily derived:
\begin{equation}
    \frac{d f_\mathrm{D}}{dt} = \frac{1}{2\eta\nu}\left(1 - f_\mathrm{D}\nu\frac{d\eta}{dt}\right).
    \label{eqn:delay-rate-change-full}
\end{equation}
The simple model of Section~\ref{ssec:defn} is necessary only to describe the curvature parameter $\eta$. In a two-screen model, it may apply only to the closer screen, or the "scintillation-arc screen" which produces the fringes in the dynamic spectrum and equivalently the parabola in the secondary spectrum.

\subsection{Screen locations}
It is possible to obtain
limits on the screen locations even with the single beaseline measurement, but
under certain assumptions that may be contrived. For example, one would have to
assume that the screen makes a negligible contribution to the effective
velocity. Implicitly, this would considerably bias the distance of a very nearby
screen while screens closer to the pulsar are more immune to the
approximation. We would hence defer any estimates of the location(s) of the
scattering screens pending a full 2D scattering measurement.

In addition, it is unknown from these measurements if the
$\eta=0.76\ \mathrm{s}^3$ is the same screen at 125 pc \citep{Wucknitz2019} that
causes the echo. More observations are required to establish this independently.

\subsection{Scintillation pattern arrival time delay}
The scintillation pattern arrival delay is a standalone measurement that holds
without recourse to any restrictive assumptions about scattering geometry \citep[e.g.][]{Slee1974}. Especially, we
note that there arises no need to invoke anisotropic scattering, as the minimum assumption required to obtain the delay is a frozen screen approximation \citep[see e.g.][for the algebra]{Smirnova2014}, where
the diffraction pattern due to the scattering screen is fixed in space sampled
by the motion of the observer in an appropriate reference frame. 

\section{IMPLICATIONS}\label{sec:implications}

The main results of this paper are threefold: (i) the curvature of the parabolic
arc of $\eta = 0.76\ \mathrm{s}^3$, (ii) the time delay of arrival of the
scintillation pattern at the two telescopes of $\Delta t_\mathrm{d} = 45.3$ s and (iii) the low scintillation
correlation of $\rho=0.22$ on the long baseline. We discuss their implications under the
assumptions stated above.

\begin{figure}
        \centering
        \includegraphics[scale=0.565]{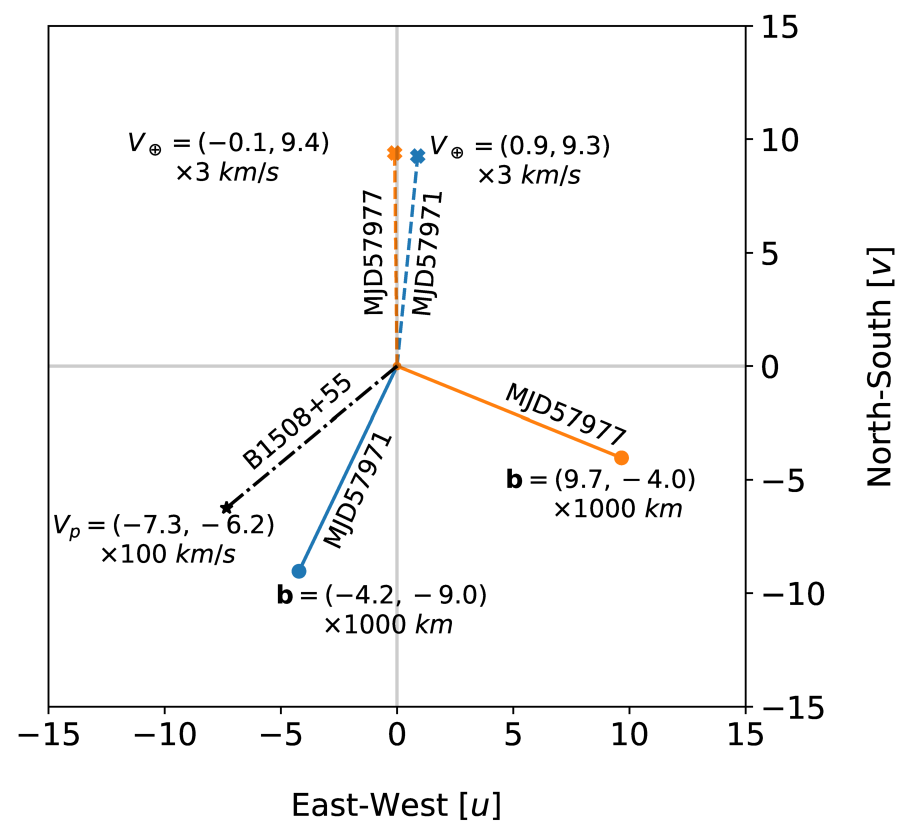}
        \caption{The orientation of the pulsar's proper motion vector, the
          Earth's motion vectors and the baseline vectors on MJD57971 and
          MJD57977 in the $u$ and $v$ directions.}
        \label{fig:all-vectors}
\end{figure}

Figure~\ref{fig:all-vectors} shows the orientation of the GMRT-ARO baseline on
MJD57971 and MJD57977 as solid lines, along with the velocity vectors of the
Earth's motion rotated to the $uvw$ co-ordinates; only the $u,v$ components are
shown with dotted lines. The $w$-component is directed upwards and normal to the $uv$ plane. The components of the pulsar's velocity in the right
ascension and declination directions are computed from the proper motion for the
distance $d_\mathrm{psr}$=2.1 kpc \citep{Chatterjee2009}.

\subsection{Multiple scattering screens: possible scenarios}\label{ssec:distances-and-velocities}

One of two possibilities
could be considered: (1) the measurement of $\Delta t_\mathrm{d} = 45.3$ s, or
the equivalent
$V^\mathrm{app}_\mathrm{ISS}=220$ km s$^{-1}$, is not associated with the $\eta =
0.76$ s$^3$ parabola at all, hinting at
the presence of a second scattering screen, (2) if instead it is indeed
associated with the $\eta = 0.76$ s$^3$ parabola, 
scattering beyond 1 $\mu$s is possibly resolved out by the baseline. 

We take note of an independent measurement of the distance to a scattering
screen using LOFAR. These VLBI observations were carried out to map the pulsed echoes 
approaching the main pulse. Direct VLBI imaging of these echoes has revealed that
they are aligned along the pulsar's proper motion vector \citep{Wucknitz2019},
i.e. $\alpha_s \sim 0^\circ$ is a reasonable approximation. They also measure
the distance to the screen of $d_\mathrm{scr}\sim125$ pc. While these pulses
were detected in the LOFAR observations, we do not detect them in the GMRT
550-750 MHz observations.
Additionally, \citet{Bansal2020} measure a distance of $\sim 251$ pc to a lens
based on their observations of B1508+55 at frequencies < 100 MHz.

\subsubsection{Is the scintillation delay associated with the LOFAR echo screen?}
\begin{figure}
    \centering
    \includegraphics[scale=0.42]{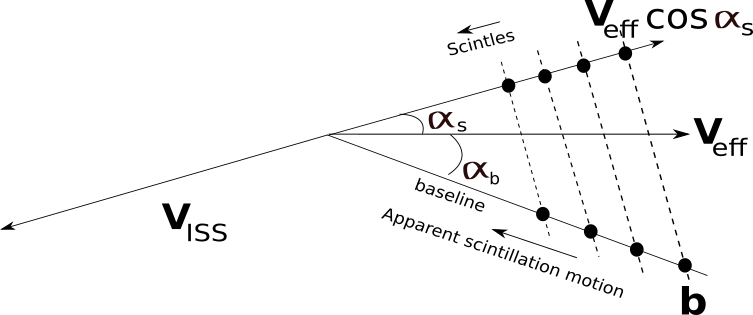}
    \caption{The pulsar proper motion vector, the scintillation
      pattern motion vector and their relative alignment with the baseline vector
      $\mathbf{b}$. The speed of scintillation pattern $V^\mathrm{app}_\mathrm{ISS}$ measured
      on the baseline is the effective velocity component along the scattering axis
      $\mathbf{V}_\mathrm{eff}\ \!\mathrm{cos} \alpha_s$ enhanced by
      $\mathrm{sec} (\alpha_\mathbf{b} - \alpha_s$). The effective velocity vector
      $\mathbf{V}_\mathrm{eff}$ is directed along the pulsar proper motion.}
    \label{fig:PM-alignment}
\end{figure}

For $d_\mathrm{scr} = 125$~pc \citep{Wucknitz2019}, the effective velocity of
scintillation from such a screen, if it is associated with the
$\eta = 0.76$ s$^3$ parabola, is $|\mathbf{V}_\mathrm{eff}| \sim47$
km s$^{-1}$. If the apparent scintillation pattern velocity of $V_\mathrm{ISS}^\mathrm{app}$ = 220 $\mathrm{km\ s^{-1}}$ is associated with the echo, then $\alpha_s = 0^\circ$, and its effective velocity should be $V_\mathrm{ISS}^\mathrm{app} \mathrm{cos} \alpha_\mathrm{\bf b}$ = 200 km s$^{-1}$ (see Figure~\ref{fig:PM-alignment}). 

If the
LOFAR measurement is therefore associated with the $\eta = 0.76$~s$^3$ parabola,
it certainly hints at the presence of
a second screen, whose location is unknown but likely closer to the
observer than the pulsar, beyond the 125-pc screen and not
associated with the  $\eta = 0.76$~s$^3$ parabola. In that case,
it is this second screen we detect on the MJD57971 baseline. The scintillation delay measured on the GMRT-ARO baseline is hence \emph{not} associated with the LOFAR echo. However, we cannot conclusively establish, but
can only propose, that the 125-pc LOFAR echo screen causes the $\eta = 0.76$ s$^3$ parabola
in the GMRT secondary spectra. 

\subsubsection{Is the LOFAR echo screen associated with the $\eta = 0.76$ s$^3$ parabola?}
Let us now consider the LOFAR measurement on its own. We
assume the 125-pc screen is practically at rest with respect to the
solar system barycentre. The pulsar's contribution to the inferred
effective velocity ($\sim47$ km s$^{-1}$, from eqn.~\ref{eqn:curvature}) is $\sim62$ km s$^{-1}$, to which
the Earth's orbital velocity adds between -30 and 30 km s$^{-1}$ by way of
annual modulation. It is hence
not unreasonable to propose that, if there are only two scattering
screens, the closer one at $\sim125$~pc is indeed the $\eta = 0.76$~s$^3$
screen, as the inferred effective velocity is within the range permitted by the annual modulation. The structures causing the LOFAR
echoes are then likely to be anomalous features in the scattering screen. If true, it is not surprising that we detect no echoes at 650 MHz, as
the deflection angle may yet be insufficient at the observing epoch. The spectral index is expected to be an aposteriori effect. 

\subsection{Baseline lengths, scintillation times and decorrelation}

The early spaced-receiver experiments, such as those by \citet{Slee1968} and
\cite{Rickett1973}, met with mixed success: either they detected no measurable delay
between the scintillation arrival on a 325-km baseline \citep{Slee1968} or measured no
decorrelation on a 5500-km baseline \citep{Lang1970}. \citet{Rickett1973}, however,
detect considerable decorrelation on a 5000-km baseline for PSR B1919+21, concluding
that the characteristic scintillation scale is smaller than the baseline. \citet{Slee1974} infer scintillation scales smaller than their 8000-km baseline. On the other
extreme, \citet{Gwinn2016} note that for the very long earth-space baselines (60,000-235,000 km; see also \citealt{Smirnova2014, Popov2017}) they observed with, the scintillation dynamic spectra should fully 
decorrelate. Our measured cross-correlation coefficient of 0.22 and a $\sim$45s pattern arrival delay on a 10,000 km baseline are between the two extremes.
For fully isotropic scattering, one can estimate the the full width at half-maximum \citep[eqn. 1][eqn. 7]{Britton1998, Popov2017} of a Gaussian scattering disc as
\begin{equation}
    \theta_\mathrm{H} = \frac{1}{\pi U_\lambda} \left[ 2\ \mathrm{ln} 2\ .\ \mathrm{ln}\frac{1}{\rho^2}\right]^{\frac{1}{2}}
\end{equation}
where $\rho$ is the correlation coefficient and $U_\lambda$ is the projected baseline length in wavelength units, giving $\theta_\mathrm{H}$ = 6.2 mas.

If we consider the \cite{Brisken2010} observations but with the longer GMRT-ARO baseline, they would have found nearly 100\% correlation. The highly anisotropic scattering seen in B0834+06 is bound to produce a high degree of correlation on a 10000-km baseline, progressively decorrelating on baselines approaching Earth-space separations as the received electric fields decohere.
At 650~MHz, or twice their observing frequency, the angular extent of their $\sim$35~mas scattering would be $\sim$9~mas, similar to the 10-mas resolution provided by the GMRT-ARO baseline, and hence nearly fully correlated.

The scintillation time measured from the GMRT dynamic spectra is 132-145
s. If the scattering were isotropic or strictly 1D, the scintillation should be 
highly correlated instead of the observed $\sim$20\%, even for a 45 s pattern arrival
delay. Our measurement therefore rules out both the scenarios, suggesting highly anisotropic 2D scattering as the alternative scenario. 
Interestingly, one could still invoke 1D
scattering oriented differently in each of two screens. The observed
decorrelation is to be expected for multiple
1D scattering screens with different values for $\alpha_s$, which is degenerate in
the cross-correlation coefficient with a single 2D scattering screen.

The scintillation delay measurement on the
MJD57971 baseline confined to low delays suggests the presence of a second scattering screen. The diminished cross-correlation could hence be the result of the 125-pc screen partially resolving the scattering on the screen further away, limiting the scintillation to a delay of 1 $\mu$s.

\subsection{Physical scales and association}
At 650 MHz the 10,000-km baseline gives an angular resolution of
$\sim10$ mas. At 125 pc, this translates to a transverse physical scale of
1.3 AU. If the main parabola arises from this 125-pc screen, an angular
displacement of $\sim7.5$ mas corresponds to a delay of 9 $\mu$s. We find that 75\%
of the power in the $\eta = 0.76$~s$^3$ parabola in the secondary spectrum
is at $|\tau| < 1$ $\mu$s, or equivalently at an off-axis angular displacement
of 2.5 mas if this screen is at 125 pc, beyond which the lensed images
could be very faint and therefore sensitivity-limited.

If the $\eta = 0.76$ s$^3$ parabola is due to the 125-pc screen, its
effective velocity must vary between 30 and 90 km s$^{-1}$ due to the
Earth's orbit around the sun, translating into considerable annual
modulation of the curvature of the main parabola. Regular monitoring at one or
multiple frequencies can help independently confirm or falsify the hypothesis
that the 125-pc echoes are associated with the $\eta = 0.76$~s$^3$ parabola by measuring the annual modulation of its curvature and comparing with the prediction. Simultaneous multi-frequency VLBI observations are
useful for direct imaging measurements of speckle positions
(e.g. \citealt{Brisken2010, Simard2019a, Simard2019b}). These positions can be
tracked with a set of periodic observations, which will allow testing the predictive model of \cite{Simard2018} by comparing with actual measurements.

It is interesting to note that the $\sim50$-$2000$ km LOFAR baselines
at $\sim140$ MHz detect the 125-pc screen \citep{Wucknitz2019}, whereas the $\sim 10000$-km GMRT-ARO
baseline at 650 MHz exclusively detects a screen possibly further
beyond the 125-pc screen. Therefore, combining multiple baselines
with multi-frequency observations might enable both the screens to
be detected simultaneously. Further, such observations could help
investigate if the structure on the closer 125-pc screen might be
resolving the scattered images from the screen located further.  
It is possible that the peculiar
shape of the reverse arclets seen for B1508+55, as well as the
islands of arclets off the main parabolic curve are indicative of the
same, but further insight can be gained from simulations. For example, \cite{Liu2016} argue
for a two-screen effect in their analysis and modelling of the data of
\cite{Brisken2010} for B0834+06, where the 1-ms island, which is off the main parabola, is the result of double
refraction. In the picture they propose, the images corresponding to the
1-ms island are the result of a single caustic in a second screen closer to the
observer refracting a group of speckles from the main screen. 
It is instructive to note that in the specific case of B0834+06, the structure that
causes the main parabola dominates at low delays. However for B1508+55, it is
possible that the second screen that we detect on the baseline dominates at low
delays as a result of being partially resolved by the 125~pc intervening screen, a
scenario very different from what is seen for B0834+06. 

The Galactic coordinates of B1508+55 are
$l=91.3^\circ$ and $b=52.3^\circ$: while it is interesting to note that
the edge of the local hot bubble in the direction of
B1508+55 lies at $\sim100$-$150$ pc from the sun (see \citealt{Liu2017}),
other local ionizing sources may also be present. \cite{Wucknitz2019} notes the
presence of an A2 star at 125 pc with a 1.37-pc offset from the line of
sight to the pulsar but aligned with the line through the echoes.

\section{SUMMARY AND CONCLUSIONS}

We have reported the findings based on simultaneous wideband observations of the scintillation of B1508+55 with the GMRT and ARO, centered at 650 MHz. We observed in two sessions $\sim150$ hours apart with approximately orthogonal baselines to sample the scattering in 2D. We detect correlated scintillation on the 9974-km baseline on MJD57971, and the correlation is significantly less than 100\%, hinting that the scattered image is 2D. The scintillation time of 135 s is $3\times$ the scintillation pattern arrival delay of 45 s, ruling out both isotropic and 1D scattering, but is evidence for highly anisotropic 2D scattering. We cannot precisely locate the screen based solely on the scintillation delay measurement on a single baseline.

The 100-200 MHz LOFAR observations measure a distance of 125 pc to the echoes, as well as $\alpha_s = 0^\circ$. We propose that the 125-pc screen gives rise to the $\eta = 0.76$ s$^3$ parabola, a scenario strongly favoured by the GMRT-ARO as well as LOFAR measurements. In that case, given that the correlated scintillation detected on the MJD57971 baseline does not align with the $\eta = 0.76 \mathrm{s}^3$ parabola, it is likely to originate in a screen beyond the 125-pc screen. While the 650-MHz, $\sim10,000$ km GMRT-ARO baseline appears to detect one screen, the 100-200 MHz, 50-2000 km, LOFAR baselines detect another. We have confirmed that these are necessarily two different screens. Multiple baselines spanning 100-10000 km at multiple frequencies could possibly detect both screens simultaneously, making a strong case for a multi-epoch, multi-frequency global long baseline campaign.

Finally, we advance the possibility that the peculiar shape of the reverse arclets along the main parabola is the result of the closer 125-pc screen partially resolving the scattered images of the screen beyond.

Let there be two screens, each scattering only weakly. The second screen, or the scintillation-arc screen would still see a coherent field, i.e. it would produce fringes irrespective of the first screen. In effect, each screen would produce its own parabolic arc, with the net field being the convolution of the two. If the second screen scatters strongly, one of the parabolic arcs may feature reverse arclets. However, an interesting possibility arises when the scattering roles of the screens are reversed.

Consider that the screen closer to the pulsar scatters strongly, either  isotropically or anisotropically, producing a strong scintillation pattern on the screen further from the pulsar. For appropriate scattering scales, the first screen will project variations in phase and amplitude onto the second. If one uses the simple model of Section~\ref{ssec:defn} for only the second screen, these variations will change the weight and phase of the stationary-phase points, thus shifting the positions and amplitudes of the resulting fringes in the observer plane with frequency and time, thereby ruining their coherence in the dynamic spectrum. This will have the effect of reducing the correlation on the long GMRT-ARO baseline. It will also have the effect of thickening the arc in the secondary spectrum. Quantitative estimates of the angular broadening and spatial scale of the scintillation that the first screen projects onto the second screen, and of the scale of the points of stationary phase on the second screen, would be most interesting but may be beyond the scope of the paper.

More data from further observations, employing multiple baselines, would be invaluable towards completely solving for both the screens at once. Such observations could also verify or falsify our hypothesis that the closer 125-pc screen partially resolves the images on the screen beyond.

\section{DATA AVAILABILITY}
The data underlying this article will be shared on reasonable request to the corresponding author.

\section*{ACKNOWLEDGEMENTS}
We thank the anonymous reviewer whose critical comments have helped greatly enhance the clarity of the contents and the presentation.
We dedicate this paper to the memory of Govind Swarup and Jean-Pi\`{e}rre Macquart. VRM thanks Olaf Wucknitz, Stefan Os\l{}owski and J-P Macquart for several insightful discussions. VRM was supported by a SOSCIP Consortium Postdoctoral Fellowship. VRM acknowledges support of the Department of Atomic Energy, Government of
India, under project no. 12-R\&D-TFR-5.02-0700. We thank the staff of
the GMRT that made these  observations possible. GMRT is run by the National
Centre for Radio Astrophysics of the Tata Institute of Fundamental Research. We thank the research collaboration and funding support of
Thoth Technology, Inc., who owns and operates ARO and contributed significantly
to this research. We also acknowledge the Ontario Research Fund - Research
Excellence program (ORF-RE) and NSERC. Computations were performed on the
Niagara supercomputer at the SciNet HPC Consortium. SciNet is funded by: the
Canada Foundation for Innovation; the Government of Ontario; Ontario Research
Fund - Research Excellence; and the University of Toronto. 

%%%%%%%%%%%%%%%%%%%%%%%%%%%%%%%%%%%%%%%%%%%%%%%%%%


\bibliographystyle{mnras}
\bibliography{mylist} % if your bibtex file is called example.bib

\begin{thebibliography}{}
\makeatletter
\relax
\def\mn@urlcharsother{\let\do\@makeother \do\$\do\&\do\#\do\^\do\_\do\%\do\~}
\def\mn@doi{\begingroup\mn@urlcharsother \@ifnextchar [ {\mn@doi@}
  {\mn@doi@[]}}
\def\mn@doi@[#1]#2{\def\@tempa{#1}\ifx\@tempa\@empty \href
  {http://dx.doi.org/#2} {doi:#2}\else \href {http://dx.doi.org/#2} {#1}\fi
  \endgroup}
\def\mn@eprint#1#2{\mn@eprint@#1:#2::\@nil}
\def\mn@eprint@arXiv#1{\href {http://arxiv.org/abs/#1} {{\tt arXiv:#1}}}
\def\mn@eprint@dblp#1{\href {http://dblp.uni-trier.de/rec/bibtex/#1.xml}
  {dblp:#1}}
\def\mn@eprint@#1:#2:#3:#4\@nil{\def\@tempa {#1}\def\@tempb {#2}\def\@tempc
  {#3}\ifx \@tempc \@empty \let \@tempc \@tempb \let \@tempb \@tempa \fi \ifx
  \@tempb \@empty \def\@tempb {arXiv}\fi \@ifundefined
  {mn@eprint@\@tempb}{\@tempb:\@tempc}{\expandafter \expandafter \csname
  mn@eprint@\@tempb\endcsname \expandafter{\@tempc}}}

\bibitem[\protect\citeauthoryear{{Bansal}, {Taylor}, {Stovall}  \&
  {Dowell}}{{Bansal} et~al.}{2020}]{Bansal2020}
{Bansal} K.,  {Taylor} G.~B.,  {Stovall} K.,   {Dowell} J.,  2020, \mn@doi
  [\apj] {10.3847/1538-4357/ab76bc}, \href
  {https://ui.adsabs.harvard.edu/abs/2020ApJ...892...26B} {892, 26}

\bibitem[\protect\citeauthoryear{{Brisken}, {Macquart}, {Gao}, {Rickett},
  {Coles}, {Deller}, {Tingay}  \& {West}}{{Brisken} et~al.}{2010}]{Brisken2010}
{Brisken} W.~F.,  {Macquart} J.~P.,  {Gao} J.~J.,  {Rickett} B.~J.,  {Coles}
  W.~A.,  {Deller} A.~T.,  {Tingay} S.~J.,   {West} C.~J.,  2010, \mn@doi
  [\apj] {10.1088/0004-637X/708/1/232}, \href
  {https://ui.adsabs.harvard.edu/abs/2010ApJ...708..232B} {708, 232}

\bibitem[\protect\citeauthoryear{{Britton}, {Gwinn}  \& {Ojeda}}{{Britton}
  et~al.}{1998}]{Britton1998}
{Britton} M.~C.,  {Gwinn} C.~R.,   {Ojeda} M.~J.,  1998, \mn@doi [\apjl]
  {10.1086/311427}, \href
  {https://ui.adsabs.harvard.edu/abs/1998ApJ...501L.101B} {501, L101}

\bibitem[\protect\citeauthoryear{{Chatterjee} et~al.,}{{Chatterjee}
  et~al.}{2009}]{Chatterjee2009}
{Chatterjee} S.,  et~al., 2009, \mn@doi [\apj] {10.1088/0004-637X/698/1/250},
  \href {https://ui.adsabs.harvard.edu/abs/2009ApJ...698..250C} {698, 250}

\bibitem[\protect\citeauthoryear{{Cordes}, {Rickett}, {Stinebring}  \&
  {Coles}}{{Cordes} et~al.}{2006}]{Cordes2006}
{Cordes} J.~M.,  {Rickett} B.~J.,  {Stinebring} D.~R.,   {Coles} W.~A.,  2006,
  \mn@doi [\apj] {10.1086/498332}, \href
  {https://ui.adsabs.harvard.edu/abs/2006ApJ...637..346C} {637, 346}

\bibitem[\protect\citeauthoryear{{Gupta} et~al.,}{{Gupta}
  et~al.}{2017}]{Gupta2017}
{Gupta} Y.,  et~al., 2017, \mn@doi [Current Science]
  {10.18520/cs/v113/i04/707-714}, 113, 707

\bibitem[\protect\citeauthoryear{{Gwinn}}{{Gwinn}}{2019}]{Gwinn2019a}
{Gwinn} C.~R.,  2019, \mn@doi [\mnras] {10.1093/mnras/stz894}, \href
  {https://ui.adsabs.harvard.edu/abs/2019MNRAS.486.2809G} {486, 2809}

\bibitem[\protect\citeauthoryear{{Gwinn} et~al.,}{{Gwinn}
  et~al.}{2016}]{Gwinn2016}
{Gwinn} C.~R.,  et~al., 2016, \mn@doi [\apj] {10.3847/0004-637X/822/2/96},
  \href {https://ui.adsabs.harvard.edu/abs/2016ApJ...822...96G} {822, 96}

\bibitem[\protect\citeauthoryear{{Hill}, {Stinebring}, {Barnor}, {Berwick}  \&
  {Webber}}{{Hill} et~al.}{2003}]{Hill2003}
{Hill} A.~S.,  {Stinebring} D.~R.,  {Barnor} H.~A.,  {Berwick} D.~E.,
  {Webber} A.~B.,  2003, \mn@doi [\apj] {10.1086/379191}, \href
  {https://ui.adsabs.harvard.edu/abs/2003ApJ...599..457H} {599, 457}

\bibitem[\protect\citeauthoryear{{Hill}, {Stinebring}, {Asplund}, {Berwick},
  {Everett}  \& {Hinkel}}{{Hill} et~al.}{2005}]{Hill2005}
{Hill} A.~S.,  {Stinebring} D.~R.,  {Asplund} C.~T.,  {Berwick} D.~E.,
  {Everett} W.~B.,   {Hinkel} N.~R.,  2005, \mn@doi [\apjl] {10.1086/428347},
  \href {https://ui.adsabs.harvard.edu/abs/2005ApJ...619L.171H} {619, L171}

\bibitem[\protect\citeauthoryear{{Johnson} \& {Gwinn}}{{Johnson} \&
  {Gwinn}}{2012}]{Johnson2012}
{Johnson} M.~D.,  {Gwinn} C.~R.,  2012, \mn@doi [\apj]
  {10.1088/0004-637X/755/2/179}, \href
  {https://ui.adsabs.harvard.edu/abs/2012ApJ...755..179J} {755, 179}

\bibitem[\protect\citeauthoryear{{Lang} \& {Rickett}}{{Lang} \&
  {Rickett}}{1970}]{Lang1970}
{Lang} K.~R.,  {Rickett} B.~J.,  1970, \mn@doi [\nat] {10.1038/225528a0}, \href
  {https://ui.adsabs.harvard.edu/abs/1970Natur.225..528L} {225, 528}

\bibitem[\protect\citeauthoryear{{Liu}, {Pen}, {Macquart}, {Brisken}  \&
  {Deller}}{{Liu} et~al.}{2016}]{Liu2016}
{Liu} S.,  {Pen} U.-L.,  {Macquart} J.~P.,  {Brisken} W.,   {Deller} A.,  2016,
  \mn@doi [\mnras] {10.1093/mnras/stw314}, \href
  {https://ui.adsabs.harvard.edu/abs/2016MNRAS.458.1289L} {458, 1289}

\bibitem[\protect\citeauthoryear{{Liu} et~al.,}{{Liu} et~al.}{2017}]{Liu2017}
{Liu} W.,  et~al., 2017, \mn@doi [\apj] {10.3847/1538-4357/834/1/33}, \href
  {https://ui.adsabs.harvard.edu/abs/2017ApJ...834...33L} {834, 33}

\bibitem[\protect\citeauthoryear{{Pen} \& {Levin}}{{Pen} \&
  {Levin}}{2014}]{Pen2014b}
{Pen} U.-L.,  {Levin} Y.,  2014, \mn@doi [\mnras] {10.1093/mnras/stu1020},
  \href {https://ui.adsabs.harvard.edu/abs/2014MNRAS.442.3338P} {442, 3338}

\bibitem[\protect\citeauthoryear{{Popov} et~al.,}{{Popov}
  et~al.}{2017}]{Popov2017}
{Popov} M.~V.,  et~al., 2017, \mn@doi [\mnras] {10.1093/mnras/stw2353}, \href
  {https://ui.adsabs.harvard.edu/abs/2017MNRAS.465..978P} {465, 978}

\bibitem[\protect\citeauthoryear{{Recnik}, {Bandura}, {Denman}, {Hincks},
  {Hinshaw}, {Klages}, {Pen}  \& {Vanderlinde}}{{Recnik}
  et~al.}{2015}]{Recnik2015}
{Recnik} A.,  {Bandura} K.,  {Denman} N.,  {Hincks} A.~D.,  {Hinshaw} G.,
  {Klages} P.,  {Pen} U.-L.,   {Vanderlinde} K.,  2015, arXiv e-prints, \href
  {https://ui.adsabs.harvard.edu/abs/2015arXiv150306189R} {p. arXiv:1503.06189}

\bibitem[\protect\citeauthoryear{{Reddy} et~al.,}{{Reddy}
  et~al.}{2017}]{Reddy2017}
{Reddy} S.~H.,  et~al., 2017, \mn@doi [Journal of Astronomical Instrumentation]
  {10.1142/S2251171716410117}, \href
  {https://ui.adsabs.harvard.edu/abs/2017JAI.....641011R} {6, 1641011}

\bibitem[\protect\citeauthoryear{{Rickett} \& {Lang}}{{Rickett} \&
  {Lang}}{1973}]{Rickett1973}
{Rickett} B.~J.,  {Lang} K.~R.,  1973, \mn@doi [\apj] {10.1086/152468}, \href
  {https://ui.adsabs.harvard.edu/abs/1973ApJ...185..945R} {185, 945}

\bibitem[\protect\citeauthoryear{{Simard} \& {Pen}}{{Simard} \&
  {Pen}}{2018}]{Simard2018}
{Simard} D.,  {Pen} U.-L.,  2018, \mn@doi [\mnras] {10.1093/mnras/sty1140},
  \href {https://ui.adsabs.harvard.edu/abs/2018MNRAS.478..983S} {478, 983}

\bibitem[\protect\citeauthoryear{{Simard}, {Pen}, {Marthi}  \&
  {Brisken}}{{Simard} et~al.}{2019a}]{Simard2019a}
{Simard} D.,  {Pen} U.~L.,  {Marthi} V.~R.,   {Brisken} W.,  2019a, \mn@doi
  [\mnras] {10.1093/mnras/stz2043}, \href
  {https://ui.adsabs.harvard.edu/abs/2019MNRAS.488.4952S} {488, 4952}

\bibitem[\protect\citeauthoryear{{Simard}, {Pen}, {Marthi}  \&
  {Brisken}}{{Simard} et~al.}{2019b}]{Simard2019b}
{Simard} D.,  {Pen} U.~L.,  {Marthi} V.~R.,   {Brisken} W.,  2019b, \mn@doi
  [\mnras] {10.1093/mnras/stz2046}, \href
  {https://ui.adsabs.harvard.edu/abs/2019MNRAS.488.4963S} {488, 4963}

\bibitem[\protect\citeauthoryear{{Slee}, {Komesaroff}  \& {McCulloch}}{{Slee}
  et~al.}{1968}]{Slee1968}
{Slee} O.~B.,  {Komesaroff} M.~M.,   {McCulloch} P.~M.,  1968, \mn@doi [\nat]
  {10.1038/219342a0}, \href
  {https://ui.adsabs.harvard.edu/abs/1968Natur.219..342S} {219, 342}

\bibitem[\protect\citeauthoryear{{Slee}, {Ables}, {Batchelor}, {Krishna-Mohan},
  {Venugopal}  \& {Swarup}}{{Slee} et~al.}{1974}]{Slee1974}
{Slee} O.~B.,  {Ables} J.~G.,  {Batchelor} R.~A.,  {Krishna-Mohan} S.,
  {Venugopal} V.~R.,   {Swarup} G.,  1974, \mn@doi [\mnras]
  {10.1093/mnras/167.1.31}, \href
  {https://ui.adsabs.harvard.edu/abs/1974MNRAS.167...31S} {167, 31}

\bibitem[\protect\citeauthoryear{{Smirnova} et~al.,}{{Smirnova}
  et~al.}{2014}]{Smirnova2014}
{Smirnova} T.~V.,  et~al., 2014, \mn@doi [\apj] {10.1088/0004-637X/786/2/115},
  \href {https://ui.adsabs.harvard.edu/abs/2014ApJ...786..115S} {786, 115}

\bibitem[\protect\citeauthoryear{{Stinebring}}{{Stinebring}}{2007}]{Stinebring2007}
{Stinebring} D.,  2007, in {Haverkorn} M.,  {Goss} W.~M.,  eds,  Astronomical
  Society of the Pacific Conference Series Vol. 365, SINS - Small Ionized and
  Neutral Structures in the Diffuse Interstellar Medium. p.~254

\bibitem[\protect\citeauthoryear{{Stinebring}, {McLaughlin}, {Cordes},
  {Becker}, {Goodman}, {Kramer}, {Sheckard}  \& {Smith}}{{Stinebring}
  et~al.}{2001}]{Stinebring2001}
{Stinebring} D.~R.,  {McLaughlin} M.~A.,  {Cordes} J.~M.,  {Becker} K.~M.,
  {Goodman} J.~E.~E.,  {Kramer} M.~A.,  {Sheckard} J.~L.,   {Smith} C.~T.,
  2001, \mn@doi [\apjl] {10.1086/319133}, \href
  {https://ui.adsabs.harvard.edu/abs/2001ApJ...549L..97S} {549, L97}

\bibitem[\protect\citeauthoryear{{Wucknitz}}{{Wucknitz}}{2019}]{Wucknitz2019}
{Wucknitz} O.,  2019, arXiv e-prints, \href
  {https://ui.adsabs.harvard.edu/abs/2019arXiv190411347W} {p. arXiv:1904.11347}

\makeatother
\end{thebibliography}

\bsp	% typesetting comment
\label{lastpage}
\end{document}